\documentclass[aip,jap,preprint,floatfix,tightenlines,showpacs,nofootinbib]{revtex4-1}
\usepackage{graphicx}
\usepackage{bm}
\usepackage{amsmath}

\usepackage{citesort}
\usepackage[english]{babel}
\usepackage{graphicx}
\usepackage{bm,multirow}
\usepackage{amsmath,footnote,amssymb,color,mathrsfs}

\newcommand{\ACSN}{ ACS Nano }

\newcommand{\APL}{ Appl. Phys. Lett. }
\newcommand{\APB}{ Ann. Phys. (Berlin) }

\newcommand{\IEEEN}{ IEEE T. Nanotechnol. }
\newcommand{\IEEEJQE}{ IEEE J. Quantum Elect. }
\newcommand{\IJMPB}{ Int. J. Mod. Phys. B }

\newcommand{\JAP}{ J. Appl. Phys. }

\newcommand{\JMO}{ J. Mod. Optic. }
\newcommand{\JMP}{ J. Math. Phys. }
\newcommand{\jpa}{ J. Phys. A }

\newcommand{\jpc}{ J. Phys. C }
\newcommand{\JPCC}{ J. Phys. Chem. C }
\newcommand{\JPCM}{ J. Phys. Condens. Matter }

\newcommand{\NP}{ Nature Phys. }
\newcommand{\OExp}{ Opt. Express }

\newcommand{\PAMS}{ Proc. Am. Math. Soc. }
\newcommand{\PE}{ Physica E }

\newcommand{\PRA}{ Phys. Rev. A }
\newcommand{\PRB}{ Phys. Rev. B }

\newcommand{\PRL}{ Phys. Rev. Lett. }

\newcommand{\SM}{ Superlattice. Microst. }

\newcommand{\SST}{ Semicond. Sci. Tech. }

\begin{document}
\title{Electric-Field Control of Bound States and Optical Spectrum in Window-Coupled Quantum Waveguides}
\author{O. Olendski}
\email[Electronic address: ]{oolendski@sharjah.ac.ae}
\affiliation{Department of Applied Physics and Astronomy, University of Sharjah, P.O. Box 27272 Sharjah, United Arab Emirates}
\date{\today}
\begin{abstract}
Properties of the bound states of two quantum waveguides coupled via the window of the width $s$ in their common boundary are calculated under the assumption that the transverse electric field $\pmb{\mathscr{E}}$ is applied to the structure. It is shown that the increase of the electric intensity brings closer to each other fundamental propagation thresholds of the opening and the arms. As a result, the ground state, which in the absence of the field exists at any nonzero $s$, exhibits the energy $E_0$ decrease for the growing $\mathscr{E}$ and in the high-field regime $E_0$ stays practically the same regardless of the size of the connecting region. It is predicted that the critical window widths $s_{cr_n}$, $n=1,2,\ldots$, at which new excited localized orbitals emerge, strongly depend on the transverse voltage; in particular,  the field leads to the increase in $s_{cr_n}$, and, for quite strong electric intensities, the critical width unrestrictedly diverges. This remarkable feature of the electric-field-induced switching of the bound states can be checked, for example, by the change of the optical properties of the structure when the gate voltage is applied; namely, both the oscillator strength and absorption spectrum exhibit a conspicuous maximum on their $\mathscr{E}$ dependence and turn to zero when the electric intensity reaches its critical value. Comparative analysis of the two-dimensional (2D) and 3D geometries reveals their qualitative similarity and quantitative differences.
\end{abstract}
\vskip.7in
\maketitle

\section{Introduction}\label{sec_1}
Contemporary nanophotonics aims at the efficient control of the intensity and frequency of the electromagnetic radiation by manipulating electronic and optical properties of the ultra-small man-made structures.\cite{Gaponenko1} One of the most convenient and useful tools for achieving this is an application to the quantum nanosized objects of the external electric field $\pmb{\mathscr{E}}$. For the quantum well, the transverse voltage leads to the quantum-confined Stark \cite{Miller1,Miller2,Weiner1} and Franz-Keldysh \cite{Miller3} effects whose experimental discovery and theoretical explanation sparked a huge interest in the subject that continues to attract a careful attention both from fundamental point of view as well as possible technological applications;\cite{Achtstein1,Tepliakov1,Tepliakov2} for example, electroabsorption experiments on artificial CdSe structures of different dimensionality demonstrated that a static applied voltage leads to the strong broadening of their lowest-energy heavy-hole absorption band and drastically reduces the absorption efficiency. \cite{Achtstein1} The field-induced change in the absorption is strongest for the one-dimensional (1D) structures. Theoretical analysis of field-induced broadening of electroabsorption spectra of semiconductor 2D nanorods and 1D nanoplatelets confirmed that the weaker quantum confinement inside quantum wells results in a much pronounced field impact on their absorption as compared to the quantum wires (QWs).\cite{Tepliakov1}

Restriction by the surfaces of the motion of the charged particles in one direction leads to the preferential routes along the interfaces. Quantum waveguides are promising elements of the semiconductor nanoelectronic circuitry which possess rich  and exciting physical effects that need their correct theoretical description.\cite{Exner3} Experimental discovery thirty years ago of conductance quantization in GaAs-AlGaAs heterojunctions \cite{Wees1,Wharam1} was a historic milestone in the study of the nano waveguides that confirmed a quantum mechanical nature of the charge transport in them. Relevant to our discussion, let us mention that it was argued that a transverse electric field applied to the QW with  impurity \cite{Vargiamidis1} or nonuniformity \cite{Petrov1} strongly affects its transport properties leading, in particular, to the immense modification of the longitudinal $I-V$ characteristics exemplified, e.g., by the collapse of the zero-field resonances of the current along the duct.\cite{Vargiamidis1}

In the present research, a theoretical analysis of the electronic and optical processes taking place in two window-coupled straight waveguides in a transverse electric field $\pmb{\mathscr{E}}$ is provided. Previously, coherent transport phenomena in double quantum waveguide (DQW), i.e., in a system of two ducts with the common wall that has a window which couples the channels, attracted a lot of attention\cite{Wang1,Bertoni1,Harris1,Ionicioiu1,Gilbert1,Marchi1,Snyder1,Bordone1,Reichl1,Ramamoorthy3,Abdullah1,Abdullah2,Abdullah3,Hirayama1,Shailos1,Pingue1,Bielejec1,Fischer1,Ramamoorthy1,Ramamoorthy2} with the special emphasis on its applications in quantum information technology where it can be used as a qubit;\cite{Bertoni1,Ionicioiu1} namely, calculations showed that if the electron is initially injected, say, from the left, into one wire (a situation, which, without loss of generality, can be defined as a logical state $|0\rangle$), then, upon reaching the coupling window (CW), due to interference in it, a part or the whole of the de Broglie wave can be transferred, depending on the CW length, to the other quantum trough. If, after passing the opening, the charged particle propagates in the second wire only what is a logical state $|1\rangle$, one has a realization of a quantum-NOT operation\cite{Marchi1} and a situation when the electron wave is distributed equally between the guides corresponds to a square-root-of-NOT ($\sqrt{\rm NOT}$) gate. Besides the CW size, which is the most crucial parameter in determining the switching rate, the latter can be additionally controlled by the external static perpendicular magnetic\cite{Harris1,Gilbert1,Abdullah2,Abdullah3} or longitudinal electric\cite{Gilbert1} fields; by the surface acoustic waves;\cite{Bordone1} by the optical radiation characterized by the photon number, frequency and polarization;\cite{Abdullah1,Abdullah3} and by the Coulomb-like interaction between the electrons.\cite{Reichl1,Abdullah2,Abdullah3} This was accompanied by several experiments which suggest a quantum interference in the CW.\cite{Hirayama1,Shailos1,Pingue1,Bielejec1,Fischer1,Ramamoorthy1,Ramamoorthy2} It has to be noted that the whole above-mentioned research considered {\em transport} properties of the DQW, i.e., a configuration when the electron energy was greater than the fundamental propagation threshold of the system. On the other hand, for more than twenty years it has been known from the variational analysis\cite{Exner1,Bulla1} and exact mode-matching calculations\cite{Exner1} that the opening in the common wall of the two quasi-1D (Q1D) waveguides leads to the existence of the bound state inside it with its energy being smaller than the lowest propagation threshold of the wider waveguide. The same holds true for the 3D structures too.\cite{Exner2} From the physical point of view, this phenomenon is explained by the fact that the CW creates an extra space in the DQW where the particle can dwell with its transverse momentum smaller than its lowest propagation counterpart in the arms. Number of such localized orbitals depends on the width $s$ of the window increasing with the latter. It is shown below that for either geometry the ground level survives arbitrary strong electric field whereas an existence of its excited counterparts is switched by the applied voltage: the increasing intensity $\mathscr{E}$ swipes out back into the continuum the bound orbital that was supported by the flat geometry. Explanation of this electric-field-induced localization-delocalization transition is based on the analysis of the dependence on the voltage of the propagation thresholds in the opening and the arms; namely, the difference between the two decreases at the increasing field and, accordingly, the opening at the strong enough $\mathscr{E}$ is unable to retain the corpuscle inside it since it presents too shallow well. This evolution enormously affects optical properties of the structure: it is shown below that such important characteristics as oscillator strength, photoionization cross section and linear optical absorption coefficient, have a pronounced maximum as a function of the applied voltage. Since the problem of the role of dimensionality in the processes of the light-matter interaction attracted a lot of attention recently, \cite{Achtstein1,Tepliakov1,Tepliakov2,Achtstein2,Achtstein3} qualitative similarities and quantitative differences between the Q1D and 3D DQW geometries are discussed; it is argued, in particular, that the coupling between the electromagnetic radiation and the charges is stronger for the former configuration. Concrete conditions under which the predicted phenomena can be observed experimentally in semiconductor nanostructures are pointed out; namely, for the GaAs waveguides of the width about 10 nm the optical maxima that lie in the visible part of the spectrum are achieved, depending on the dimensionality, at the experimentally accessible \cite{Miller1,Miller2,Weiner1} electric intensities $\sim10^4-10^5$ V/cm.

Models that we analyze are described in Sec.~\ref{sec_Model} together with the necessary formulation. Results and their discussion are presented in Sec.~\ref{sec_Results} and Sec.~\ref{sec_Conclusions} is devoted to conclusions.

\begin{figure}
\centering
\includegraphics[width=\columnwidth]{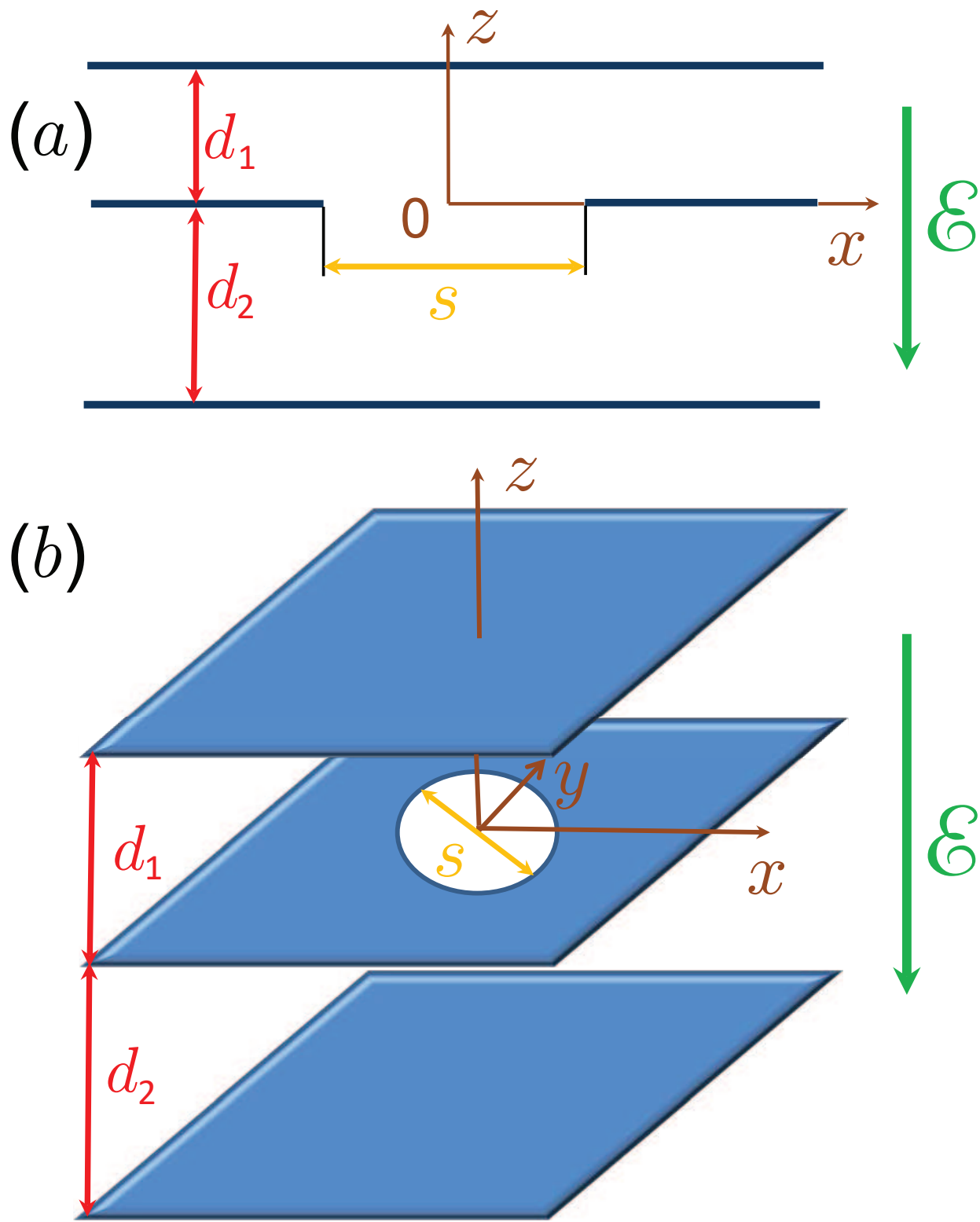}
\caption{\label{Schematics}
Two (a) Q1D and (b) 3D waveguides of generally different widths $d_1$ and $d_2$ coupled through the opening $s$ are subject to a transverse electric field $\pmb{\mathscr{E}}$.}
\end{figure}

\section{Model and Formulation}\label{sec_Model}
Structures we consider are shown schematically in Fig.~\ref{Schematics}: two straight infinitely long Q1D [panel (a)] or 3D [subplot (b)] QWs of (generally different) widths $d_1$ and $d_2$ are connected through the opening of the length (Q1D case) or diameter (3D geometry) $s$ in their common wall. In addition, normal to the interfaces electric field $\pmb{\mathscr{E}}$ is applied to the system. The first step in our analysis consists of defining the solution of the Schr\"{o}dinger equation for the wave function $\Psi({\bf r})$
\begin{equation}\label{Schrodinger1}
-\frac{\hbar^2}{2m_p}{\bm\nabla}^2\Psi({\bf r})+V({\bf r})\Psi({\bf r})-e\mathscr{E}z\Psi({\bf r})=E\Psi({\bf r})
\end{equation}
in each part of the waveguides and matching them, to find egen energy (or energies) $E$. Here, $m_p$ is a mass of the particle (for definiteness, we will talk about the electron) that, in the case of the semiconductor DQW, is an electron effective mass in the corresponding material, $e$ is the absolute value of the elementary charge, $V({\bf r})$ is electrostatic potential of the waveguides that is zero inside them and turns into infinity at the surfaces, and radius-vector $\bf r$ for the Q1D (3D) geometry is a function of the two (three) spatial coordinates: ${\bf r}=(x,z)$ [${\bf r}=(x,y,z)$]. It is convenient from the very beginning to switch to dimensionless scale where all distances are measured in units of $d\equiv\frac{d_1+d_2}{2}$, energies -- in units $\pi^2\hbar^2/(2m_pd^2)$ that is the ground-state energy of the infinitely deep Dirichlet well of the width $d$, electric intensities -- in units of $\pi^2\hbar^2/(2em_pd^3)$, electric dipole moments -- in units of $ed$, frequencies -- in units of $\pi^2\hbar/(2m_pd^2)$, and time -- in units of $2m_pd^2/(\pi^2\hbar)$. With the help of the method of separation of variables, one can write down the function $\Psi({\bf r})$ as an infinite sum of the products of the transverse and longitudinal dependencies; for example, in the 3D opening one has:
\begin{equation}\label{Psi3DRhoLessA1}
\Psi_m^<(\rho,\varphi,z)=\frac{e^{im\varphi}}{(2\pi)^{1/2}}\sum_{n=0}^\infty A_n^{|m|}Z_n^{(o)}(z)J_{|m|}\!\left(\!\pi\sqrt{E-E_n^{(o)}}\rho\right),\quad\rho\le s/2,
\end{equation}
where, due to the cylindrical symmetry, instead of the Cartesian $x$ and $y$ coordinates, their polar counterparts $\rho=(x^2+y^2)^{1/2}$ and $\varphi=\arctan(y/x)$ have been introduced. Also, $m=0,\pm1,\pm2,\ldots$ is an azimuthal quantum index, $J_\nu(\xi)$ is $\nu$th order Bessel function of the first kind,\cite{Abramowitz1} and normalized to unity,
\begin{equation}\label{NormalizationZ1}
\int_{-d_2}^{d_1}\left[Z_n^{(o)}(z)\right]^2dz=1,
\end{equation}
transverse part $Z_n^{(o)}(\mathscr{E};z)$ is expressed with the help of the Airy functions Ai$(\xi)$ and Bi$(\xi)$ \cite{Abramowitz1,Vallee1} whereas the threshold energies $E_n^{(o)}(\mathscr{E})$ are found from the requirement of vanishing of $Z_n^{(o)}(z)$ at the walls, $Z_n^{(o)}(-d_2)=Z_n^{(o)}(d_1)=0$, what for the equal widths, $d_1=d_2=1$, reduces to $Z_n^{(o)}(\pm1)=0$. Detailed analysis of the properties of $Z_n^{(o)}(z)$ and $E_n^{(o)}$ is given in Ref.~\onlinecite{Olendski1}.\footnote{Advantages of using analytic expressions for the solution in the form of the Airy functions were demonstrated on the example of the 1D quantum well with an arbitrary permutation of the Dirichlet and Neumann boundary conditions\cite{Olendski1} or Robin wall\cite{Olendski2} in the electric field when physical observables such as, for example, dipole moment are evaluated analytically too; in addition, an efficient computation with their help of the eigen energies was employed in the study of the thermodynamic properties of these structures.\cite{Olendski3,Olendski4}} In the duct region, one seeks for the solution of the form
\begin{equation}\label{Psi3DRhoGreaterA1}
\Psi_m^>(\rho,\varphi,z)=\frac{e^{im\varphi}}{(2\pi)^{1/2}}\sum_{n=0}^\infty B_n^{|m|}Z_n^{(d)}(z)K_{|m|}\!\left(\!\pi\sqrt{E_n^{(d)}-E}\rho\right),\quad\rho\geq s/2.
\end{equation}
Here, similar to the asymmetric field-free case,\cite{Exner1} $E_n^{(d)}$ form a non decreasing sequence uniting eigen energies from the lower and upper waveguides with the associated transverse functions $Z_n^{(d)}(z)$, and $K_\nu(\xi)$ is $\nu$th order modified Bessel function.\cite{Abramowitz1} For the Q1D case, the states are conveniently separated into symmetric, $\Psi(-x,z)=\Psi(x,z)$, and anti symmetric $\Psi(-x,z)=-\Psi(x,z)$, ones with respect to the reflection around the $z$ axis; for example, for the former case, one has:
\begin{subequations}\label{Functions1D}
\begin{eqnarray}\label{Functions1D_Opening1}
\Psi^<(x,z)&=&\sum_{n=0}^\infty A_nZ_n^{(o)}(z)\cos\!\left(\!\pi\sqrt{E-E_n^{(o)}}x\right),\quad |x|\le s/2,\\
\label{Functions1D_Arms1}
\Psi^>(x,z)&=&\sum_{n=0}^\infty B_nZ_n^{(d)}(z)\exp\!\left(\!-\pi\sqrt{E_n^{(d)}-E}|x|\right),\quad |x|\geq s/2.
\end{eqnarray}
\end{subequations}
Bound-state energies that satisfy inequality 
\begin{equation}\label{EnergyRequirement1}
E_0^{(o)}<E<E_0^{(d)}
\end{equation}
are found from matching the wave functions at $\rho=s/2$ or $x=s/2$. Note that due to the fact that these energies are also smaller than the first excited propagation threshold in the opening, $E<E_1^{(o)}$, Bessel functions $J_{|m|}\!\left(\!\pi\sqrt{E-E_n^{(o)}}\rho\right)$ from Eq.~\eqref{Psi3DRhoLessA1} [cosines $\cos\left(\!\pi\sqrt{E-E_n^{(o)}}x\right)$ from Eq.~\eqref{Functions1D_Opening1}] for $n\geq1$ have to be replaced by $I_{|m|}\!\left(\!\pi\sqrt{E_n^{(o)}-E}\rho\right)$ $\left[{\rm hyperbolic\thinspace cosines} \cosh\!\left(\!\pi\sqrt{E_n^{(o)}-E}x\right)\right]$, where $I_\nu(\xi)$ is another modified Bessel function.\cite{Abramowitz1} After this, the coefficients $A_n$ and $B_n$ are calculated using the fact that the total waveform has to be normalized to unity:
\begin{equation}\label{NormalizationPsi1}
\int|\Psi({\bf r})|^2dV=1.
\end{equation}
Knowledge of the energies and wave functions allows one to define all other characteristics of the system; in particular, in our analysis below a crucial role will be played by the polarization matrix:
\begin{equation}\label{Polarization1}
P_{nn'}=\left\langle\Psi_nz\Psi_{n'}\right\rangle=\int z\Psi_n^*({\bf r})\Psi_{n'}({\bf r})dV,
\end{equation}
whose diagonal elements define the electric dipole moment $p_n$ of the corresponding state:
\begin{equation}\label{Polarization2}
p_n\equiv P_{nn}=\int z|\Psi_n({\bf r})|^2dV.
\end{equation}
From computational point of view, it has to be noted that all integrals in Eqs.~\eqref{NormalizationZ1} , \eqref{NormalizationPsi1} and \eqref{Polarization2} are evaluated analytically.\cite{Vallee1,Gradshteyn1,Prudnikov1,Prudnikov2}

\begin{figure}
\centering
\includegraphics[width=\columnwidth]{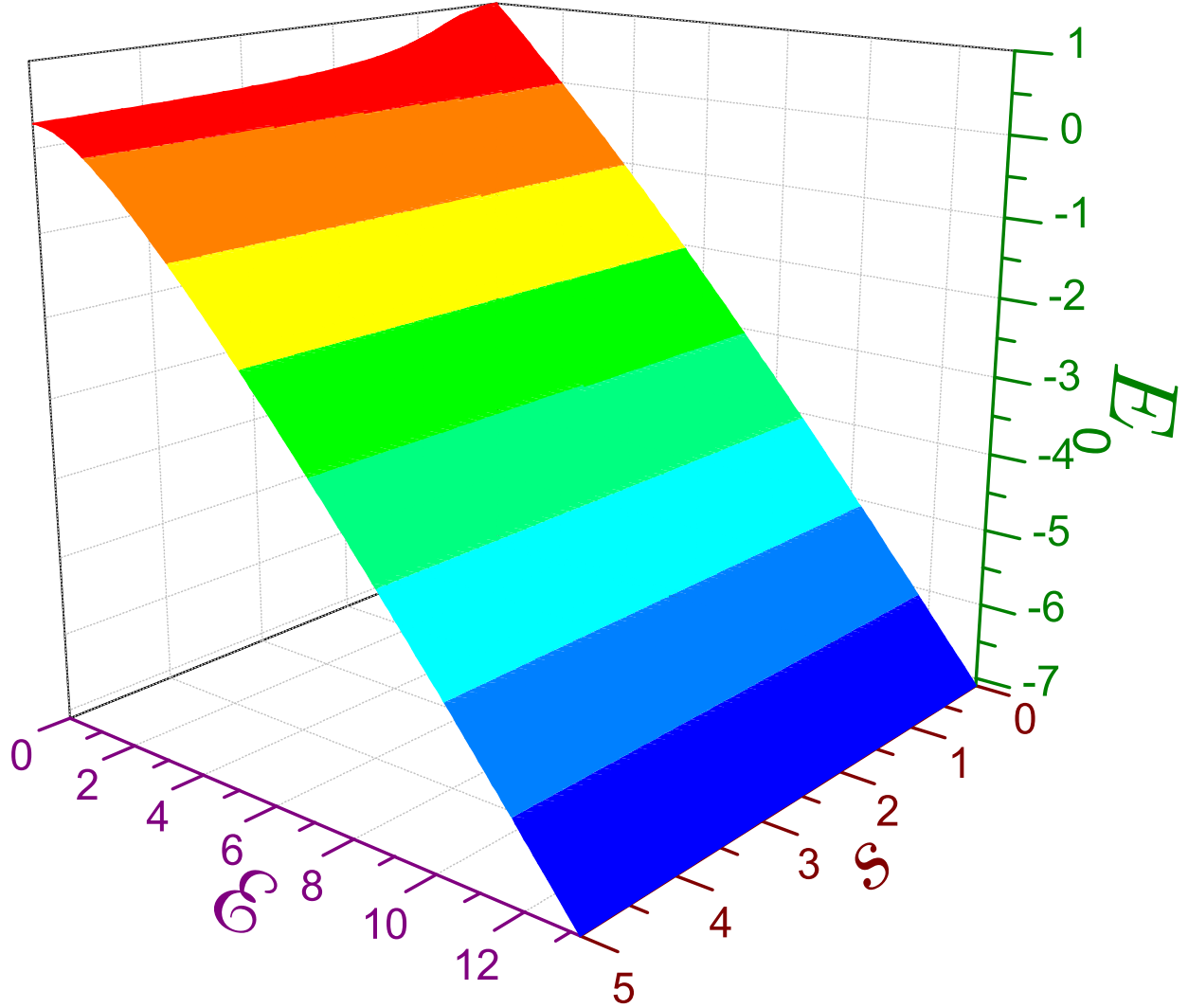}
\caption{\label{E0vsElectricAndA}
Ground-state energy of the Q1D waveguide as a function of the electric field $\mathscr{E}$ and width of the opening $s$.}
\end{figure}
\section{Results and discussion}\label{sec_Results}
\begin{figure}
\centering
\includegraphics[width=\columnwidth]{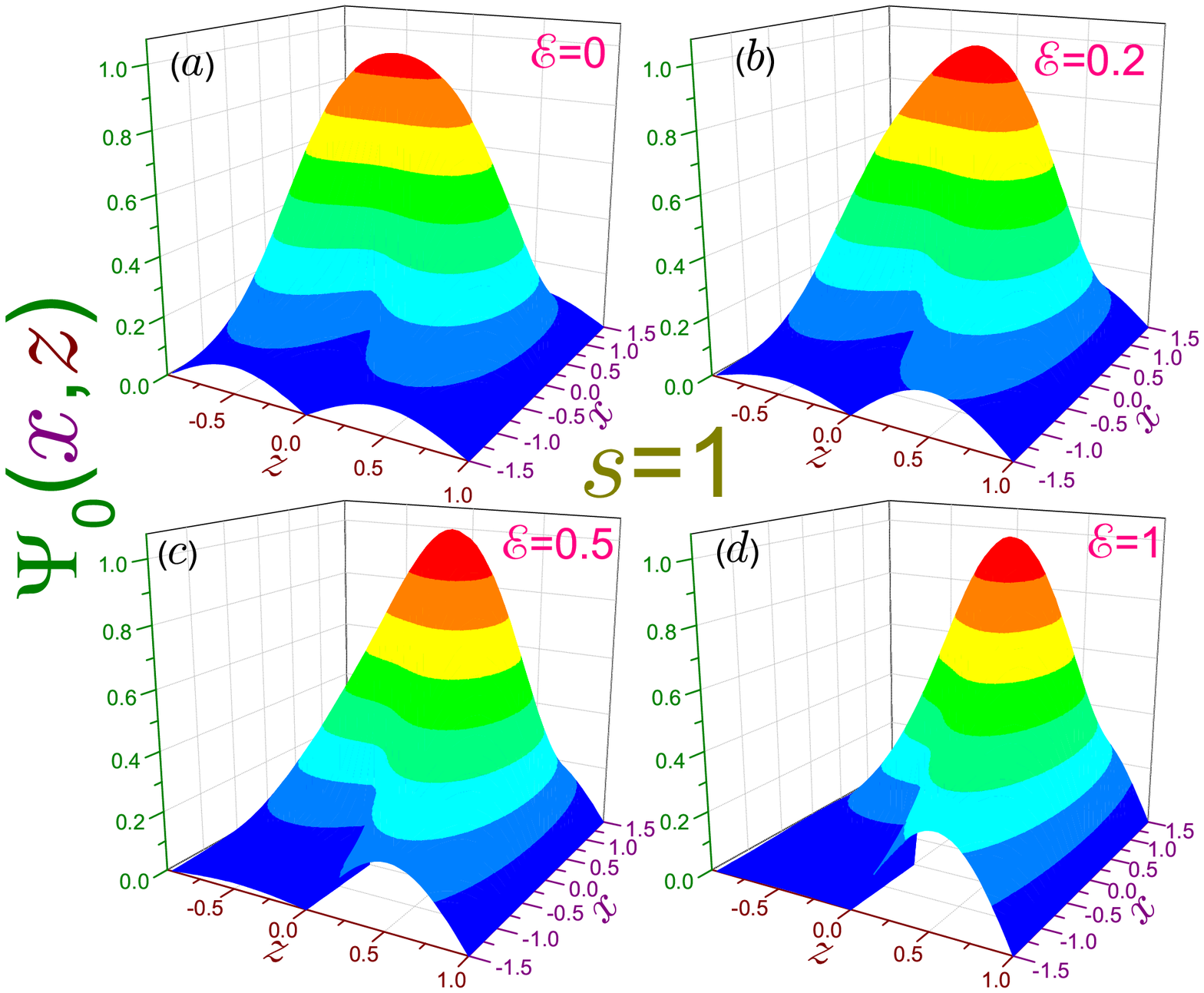}
\caption{\label{FunctionsQuasi1DFig1}
Ground-state waveform $\Psi_0(x,z)$ of the Q1D waveguide for $s=1$ and several electric fields $\mathscr{E}$ denoted in the corresponding panels.}
\end{figure}

Previous variational calculations showed that the ground state survives any electric field $\mathscr{E}$ for both geometries.\cite{Matveev1,Matveev2} Our results confirm this statement. Fig.~\ref{E0vsElectricAndA} shows Q1D ground-state energy in terms of the electric intensity $\mathscr{E}$ and size of the connecting region $s$ for the equal widths, $d_1=d_2=1$. At the zero field, the energy monotonically diminishes from unity at $s=0$ to $1/4$ at the wide openings. Applied voltage decreases the energy which at the strong fields almost does not depend on $s$. The reason for this will become clear shortly.

\begin{figure}
\centering
\includegraphics[width=\columnwidth]{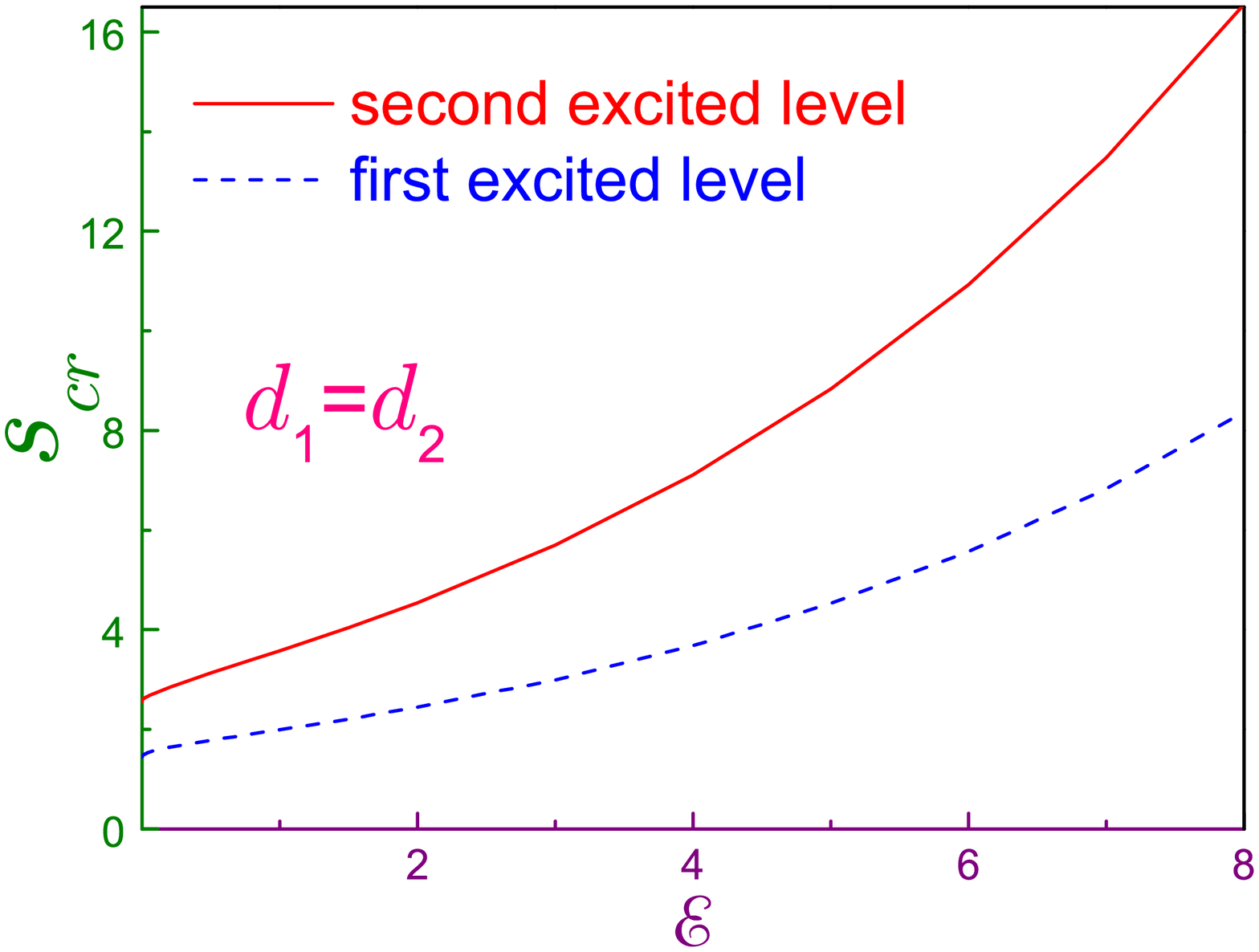}
\caption{\label{CriticalLengthFig1}
Critical widths $s_{cr}$ at which first (dashed line) and second (solid curve) excited bound state of the Q1D waveguide emerge as functions of the applied voltage.}
\end{figure}

Evolution with the electric field of the ground-state Q1D wave function is shown in Fig.~\ref{FunctionsQuasi1DFig1} for $s=1$. It is seen that for the increasing intensity $\mathscr{E}$ the electron is pushed stronger into the waveguide located at $0\leq z\leq1$; i.e., quasi classically, one can say that the particle follows the direction of the force applied to it. The transverse displacement is accompanied by the stronger longitudinal leakage of the waveform in the upper arm: there, $\Psi_0(x,z)$ at $|x|\geq s/2$ grows with the field; however, the magnitude of its sole maximum at $x=0$ stays practically intact by the applied voltage. It is instructive to compare this bound-state behavior with the scattering configuration; videlicet, as was mentioned in the Introduction, the great promise of the transport properties of the DQW is its ability to perform quantum logic gates by the controlled switching of the charge current between the ducts; for example, at very specific CW lengths that determine the intensity of the interference inside the opening, the electric flow can be switched completely from one wire to its second counterpart what realizes a quantum-NOT operation.\cite{Bertoni1,Harris1,Marchi1} The total switch of the electron after passing the CW can be also achieved at the fixed opening and very peculiar characteristics of the optical radiation\cite{Abdullah1} which serves as an additional means of controlling the NOT operation. On the contrary, the result of the electric field influence on the ground bound orbital is qualitatively the same at any arbitrary CW length: by destroying a transverse symmetry of the DQW, it monotonically increases the presence of the charge in one quantum trough simultaneously  decreasing the probability of finding the particle in the second waveguide; however, this latter probability, which becomes exponentially small at the high voltages, never turns to zero what means an impossibility of the complete switch of the localized charge in DQW.
\begin{figure}
\centering
\includegraphics[width=\columnwidth]{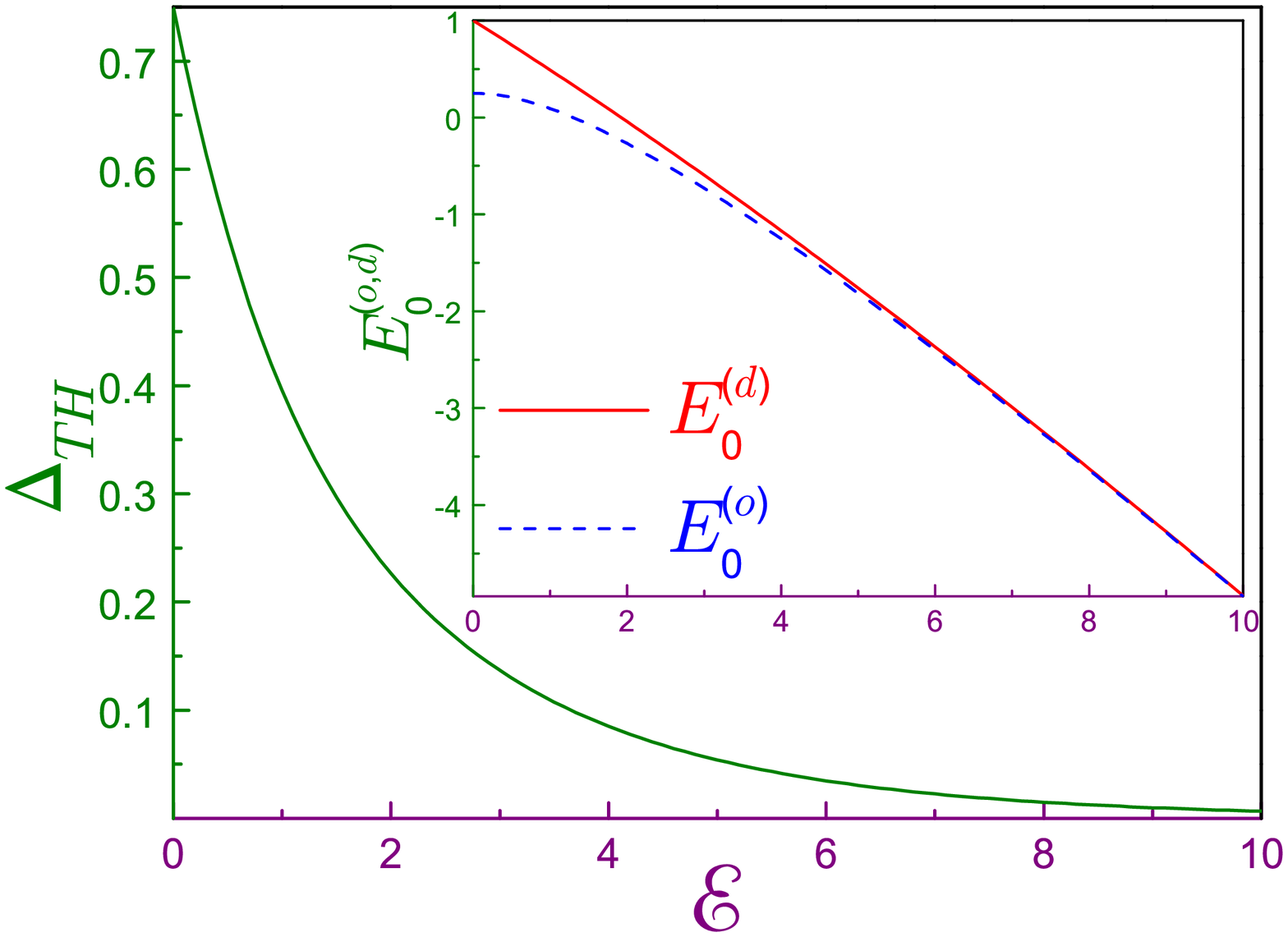}
\caption{\label{ThresholdDifferenceFig1}
Difference $\Delta_{TH}$, Eq.~\eqref{ThresholdDifference}, as a function of the applied voltage. Inset shows energies $E_0^{(d)}$ and $E_0^{(o)}$.}
\end{figure}
\begin{figure}
\centering
\includegraphics[width=\columnwidth]{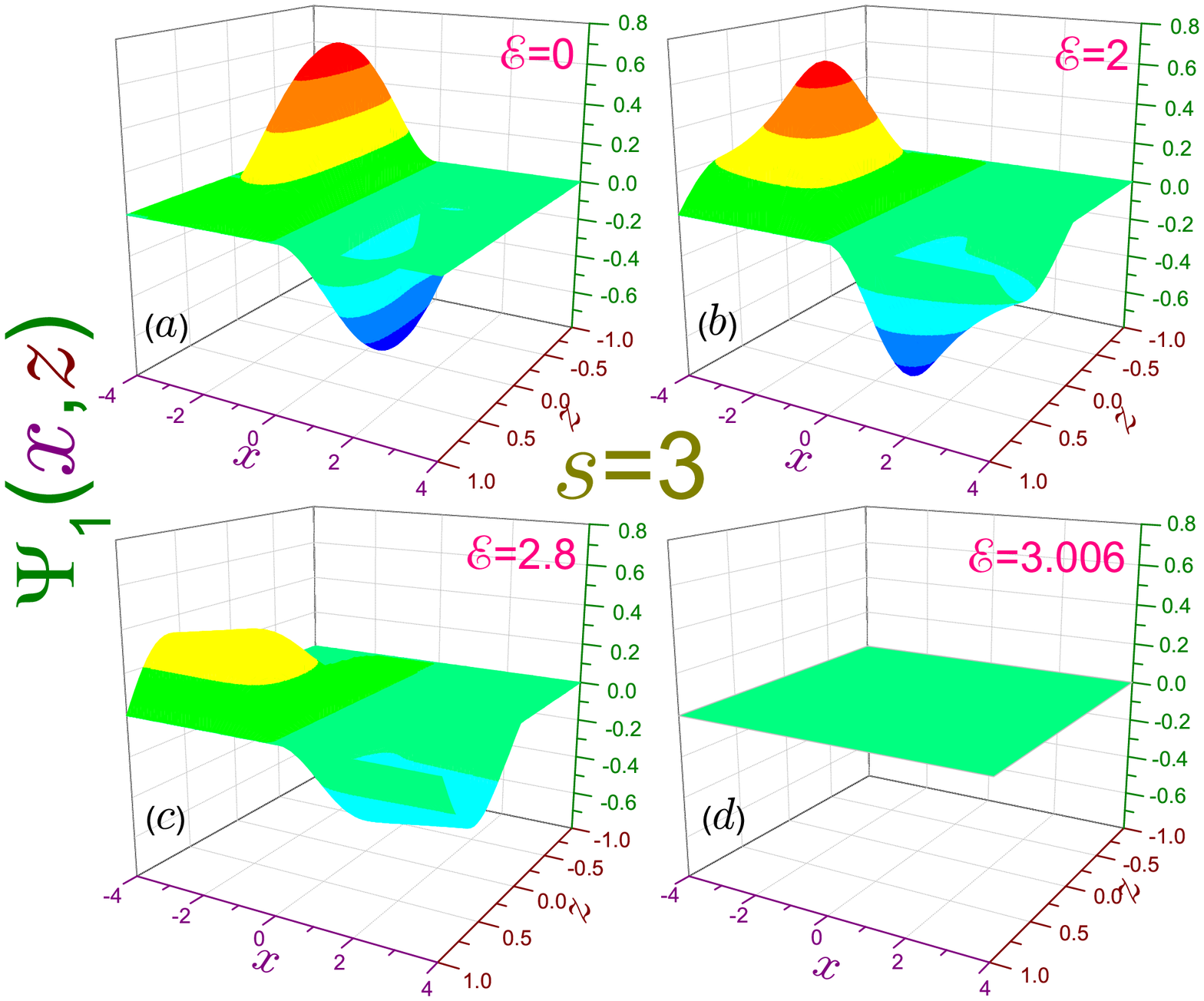}
\caption{\label{FunctionsFirstExcitedFig1}
Wave function $\Psi_1(x,z)$ of the first excited state of the Q1D waveguide for $s=3$ and several electric fields $\mathscr{E}$ denoted in the corresponding panels. Note different $\Psi$ and $x$ ranges as compared to Fig.~\ref{FunctionsQuasi1DFig1}.}
\end{figure}

It is known that for the field-free case, excited bound states exist at the wide enough opening only; for example, at the equal widths, $d_1=d_2=1$, the second Q1D localized level with its odd wavefunction emerges from the continuum at $s_{cr_1}=1.447$ whereas the next symmetric orbital is observed at the lengths greater than $s_{cr_2}=2.560$, etc.\cite{Exner1} For the 3D geometry, ground state is cylindrically symmetric, $m=0$, while the second level with $|m|=1$ exists at the diameter not smaller than $2.054$ and is followed by $|m|=2$ orbital with $s_{cr}=3.098$ and another $m=0$ state with its critical diameter being equal to $3.132$.\cite{Najar1} Applied voltage destroys a transverse symmetry of the structure leading in this way to the changes of the critical widths $s_{cr_n}$. Fig.~\ref{CriticalLengthFig1} depicts their dependencies on the intensity $\mathscr{E}$ for the two lowest excited Q1D levels. It is seen that $s_{cr_n}$ increases with the voltage. This happens since the difference $\Delta_{TH}$ between fundamental propagation thresholds in the opening and in the duct part 
\begin{equation}\label{ThresholdDifference}
\Delta_{TH}=E_0^{(d)}-E_0^{(o)}
\end{equation}
\begin{figure}
\centering
\includegraphics[width=\columnwidth]{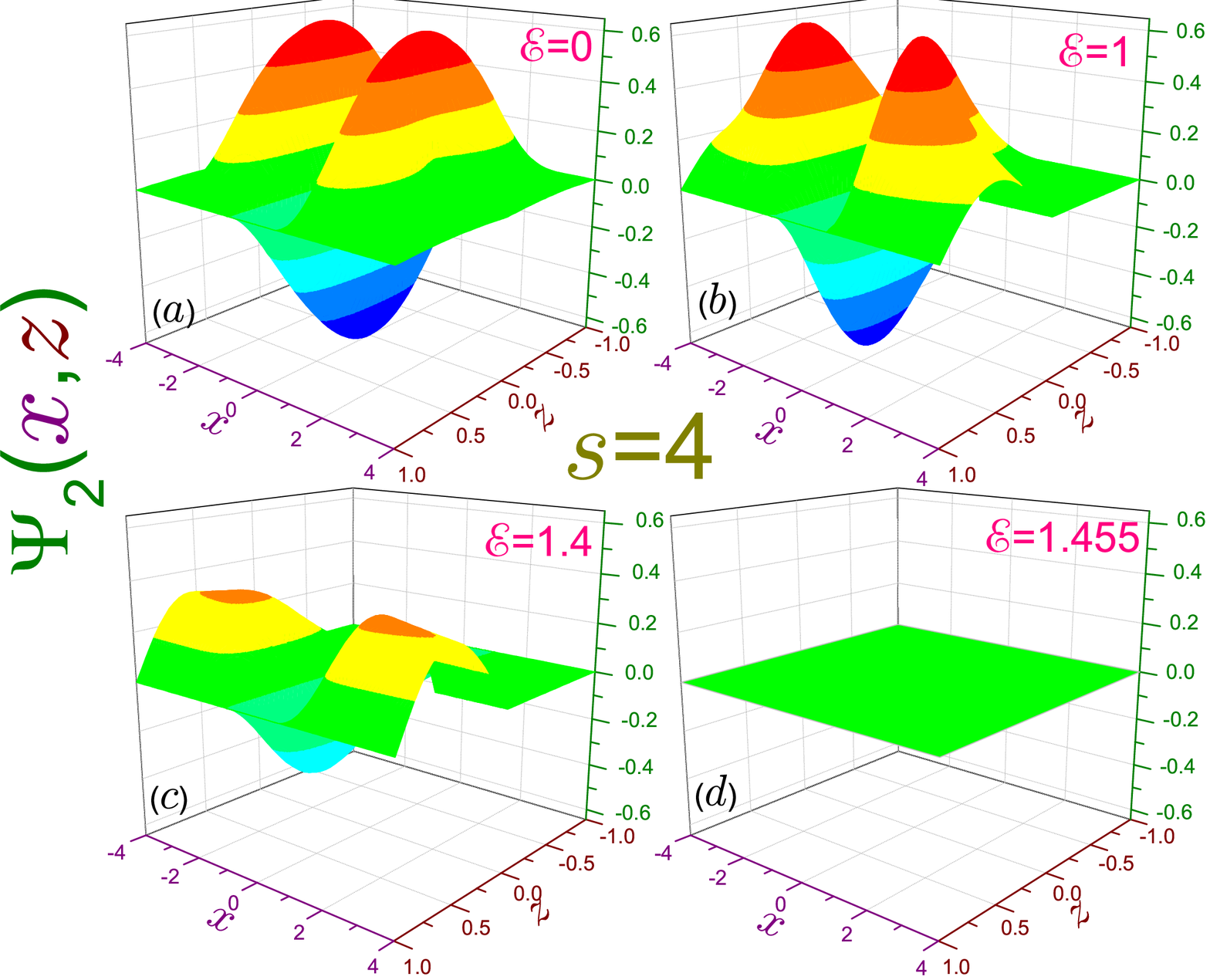}
\caption{\label{FunctionsSecondExcitedFig1}
The same as in Fig.~\ref{FunctionsFirstExcitedFig1} but for the second excited state and $s=4$. Note different $\Psi$ range as compared to Fig.~\ref{FunctionsFirstExcitedFig1}.}
\end{figure}
is a monotonically decreasing function of the field, as Fig.~\ref{ThresholdDifferenceFig1} demonstrates. Utilizing asymptotics of the Airy functions and properties of the solutions in the opening and the duct,\cite{Olendski1} it is elementary to show that at the strong voltage and equal widths the propagation thresholds in the opening and the upper duct are, respectively:
\begin{subequations}\label{AsymptoticThresholds1}
\begin{eqnarray}\label{AsymptoticThresholds1_Opening1}
E_n^{(o)}&=&-\left(\frac{\mathscr{E}}{\pi}\right)^{2/3}\!\!a_n-\mathscr{E}+\frac{1}{2\pi}\frac{1}{{\rm Ai}'(a_n)^2}\left(\frac{\mathscr{E}}{\pi}\right)^{2/3}\exp\!\left(-2^{3/2}\frac{4}{3}\pi\mathscr{E}\right)\\
\label{AsymptoticThresholds1_Duct1}
E_n^{(d_{up})}&=&-\left(\frac{\mathscr{E}}{\pi}\right)^{2/3}\!\!a_n-\mathscr{E}+\frac{1}{2\pi}\frac{1}{{\rm Ai}'(a_n)^2}\left(\frac{\mathscr{E}}{\pi}\right)^{2/3}\exp\!\left(-\frac{4}{3}\pi\mathscr{E}\right),
\end{eqnarray}
\end{subequations}
$\mathscr{E}\gg1$. For $d_1=d_2=1$, propagation energies in the lower duct $E_n^{(d_{low})}$ are always expressed as $E_n^{(d_{low})}=E_n^{(d_{up})}+\mathscr{E}$, and a proper amalgamation of this subset with $E_n^{(d_{up})}$, as stated above, forms the infinite sequence $E_n^{(d)}$ that is arranged in a non decreasing order and whose functions $Z_n^{(d)}(z)$ present a complete set at the interval $[-1,1]$. In Eqs.~\eqref{AsymptoticThresholds1}, $a_n$ is the $n$th root of the Airy function, ${\rm Ai}(a_n)=0$.\cite{Abramowitz1,Vallee1} They show that in the high-voltage regime the difference $\Delta_{TH}$ is getting negligibly small: it is determined  as a subtraction of two fading exponents. Physically, it is explained by the fact that at the strong fields the particle is pushed to the one wall only and, accordingly, it almost does not 'feel' the second surface irrespectively of the distance to it. Convergence of the energies $E_0^{(d)}$ and $E_0^{(o)}$ with the growing field is demonstrated in the inset of Fig.~\ref{ThresholdDifferenceFig1}. Due to disappearing difference $\Delta_{TH}$, the ground state energy at high electric intensities almost does not depend on the length $s$, as it was noted above during discussion of Fig.~\ref{E0vsElectricAndA}. Variation of the critical width with the field also means that the applied voltage can switch localized states in the structure; namely, for the opening $s$ that is longer than $s_{cr_n}(\mathscr{E})$, the bound orbital does exist but the increasing electric intensity destroys it pushing it back into the continuum. For each $n\geq1$, the corresponding line in Fig.~\ref{CriticalLengthFig1} divides the $s-\mathscr{E}$ plane into two parts: above it a localized orbital does exist whereas below it the corresponding state is distributed uniformly all over the duct without square integrable function $\Psi$. It is very illuminating to draw parallels to the 1D quantum well;\cite{Gaponenko1,Landau1} namely,  very shallow symmetric well with the vanishingly small difference between its bottom and the top (what in our case corresponds to the high-field regime) always has at least one bound level with its almost width-independent energy near the top. If its extent is too narrow, the well is too 'weak' to support other localized states; however, elongating (for our geometry, it means increasing the opening $s$) or deepening (what for the case of the coupled waveguides corresponds to the decrease of the applied voltage, Fig.~\ref{ThresholdDifferenceFig1}) it, one is able to capture more spatially confined orbitals. A strong similarity between the two configurations is clearly seen.

Evolution of the waveforms for the first and second excited states is shown in Figs.~\ref{FunctionsFirstExcitedFig1} and \ref{FunctionsSecondExcitedFig1}, respectively. It is seen that, contrary to the ground level, the growing field decreases the magnitudes of the extrema of the functions and at the critical electric intensity they become perfectly flat with zero magnitude.

\begin{figure}
\centering
\includegraphics[width=0.87\columnwidth]{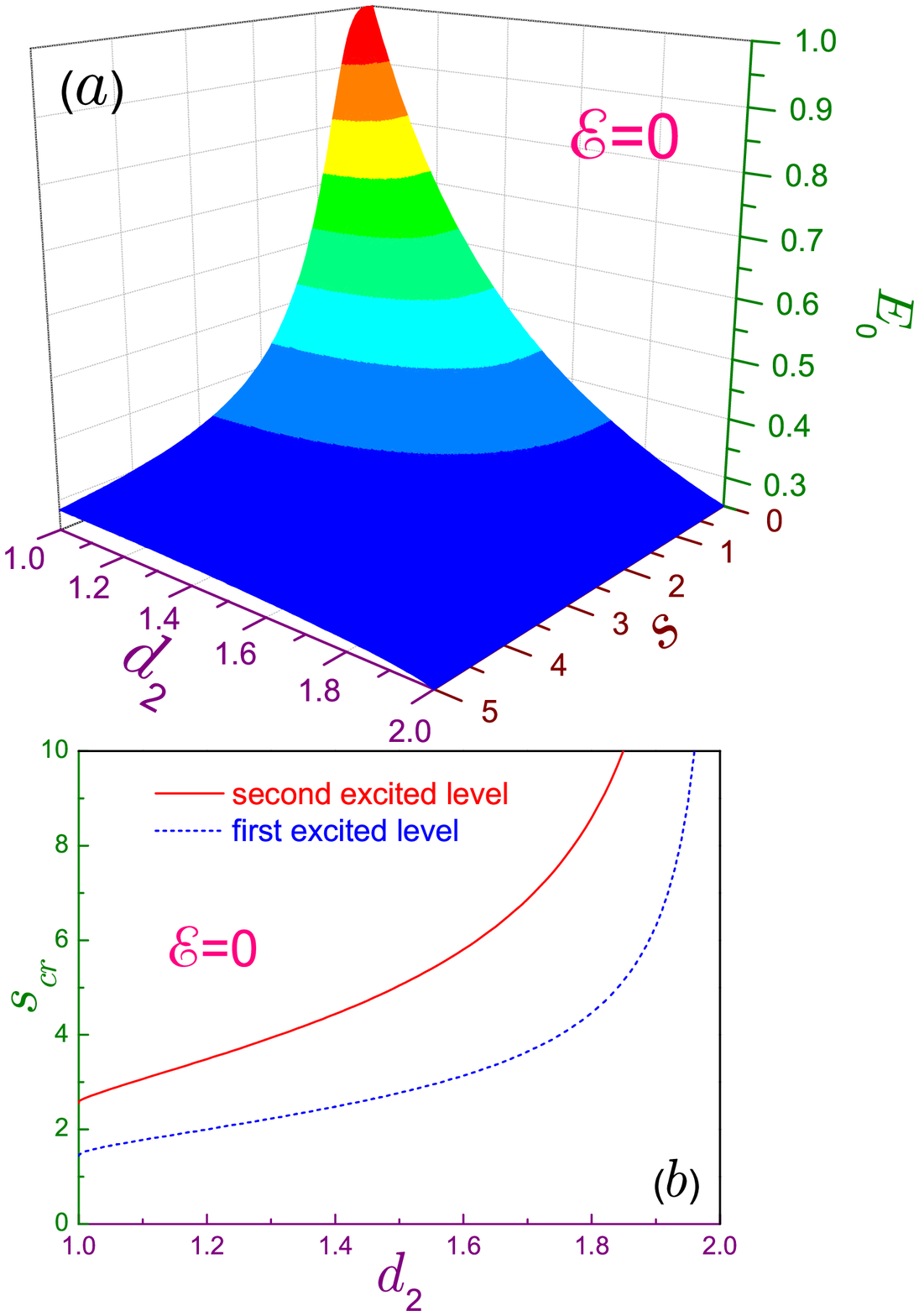}
\caption{\label{AcrVersusD2Fig1}
(a) Bound-state energy of the flat DQW, as a function of the width $d_2$ and the opening $s$. (b) Critical widths $s_{cr}$ for the two lowest excited levels as functions of $d_2$.}
\end{figure}

Above, a situation was described when a  transverse symmetry of the structure with equal duct widths, $d_1=d_2=1$, is destroyed by the applied voltage. In addition, let us point out that the flat DQW, $\mathscr{E}=0$, is also asymmetric when the wire widths are different from each other, $d_1\neq d_2$. A comparison of these two transverse-symmetry-breaking geometries reveals a lot of similarities between them; namely, in the latter configuration, the ground orbital exists at any $1\leq d_2<2$,\cite{Exner1} as it is the case for the nonzero bias and equal widths, and if the deviation from the symmetry is very large (what means that one of the waveguides is very narrow, $d_1\ll1$), its energy almost does not depend on the opening size since the propagation thresholds in both parts of the system are about the same and equal to $1/4$.  This directly corresponds to the strong-field situation, $\mathscr{E}\gg1$, when its growing magnitude makes the transverse asymmetry stronger resulting in the independence of the ground-state energy on CW length, as discussed above. Panel (a) of Fig.~\ref{AcrVersusD2Fig1} that shows ground-state energy dependence on $d_2$ and $s$ exemplifies this behavior. Moreover, stronger departure from the symmetry results in the growth of the critical widths of the excited levels, and in the extremely asymmetric geometry, $d_2\rightarrow2$, they diverge, as Fig.~\ref{AcrVersusD2Fig1}(b) demonstrates. This draws straight parallels to the high-voltage configuration when $s_{cr_n}$ unrestrictedly increase with $\mathscr{E}$ tending to infinity, Fig.~\ref{CriticalLengthFig1}. In either case, the physical reason of these divergences lies in the decrease of the difference between propagation thresholds in the arms and the opening with the growth of the transverse asymmetry. It has to be noted that  inevitable structural asymmetry appears quite naturally when split-gate method is used for creating complicated nanostructures\cite{Ramamoorthy2} and since for quantum computing applications it is a significant detrimental factor, special measures have to be implemented to eliminate the difference between the constituent wires.\cite{Ramamoorthy2}

\begin{figure}
\centering
\includegraphics[width=\columnwidth]{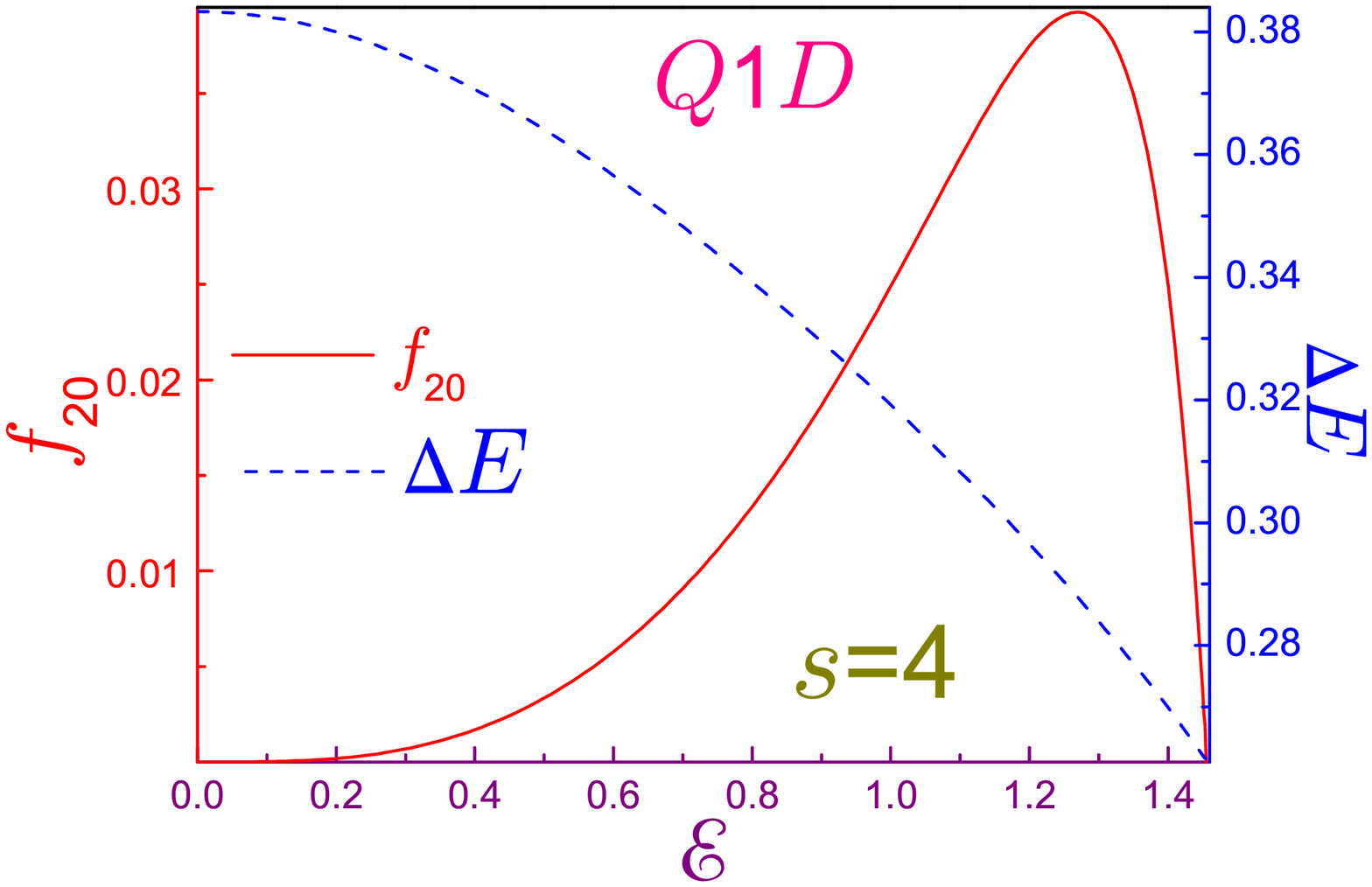}
\caption{\label{OscillatorStrengthQuasi1DFig1}
Oscillator strength $f_{20}$ of the Q1D waveguide with $s=4$ (solid line, left axis) and corresponding energy difference $\Delta E$ (dashed curve, right axis) as a function of the electric field $\mathscr{E}$.}
\end{figure}

Having seen the evolution of the bound states in the electric fields, one should wonder whether it can be observed experimentally. To answer affirmatively to this question, let us show that the above described behavior of the excited localized levels has its drastic consequences on the optical properties of the waveguide. Consider, for example, a monochromatic field of frequency $\omega$ that illuminates the waveguide with its electric component perpendicular to the interfaces:
\begin{equation}\label{Light1}
\pmb{\mathscr{E}}_{opt}(z,t)=\mathscr{E}_0ze^{i\omega t}{\bf k},
\end{equation}
where $\bf k$ is a unit vector along $z$ axis. Treating this linearly polarized light as a small perturbation, one derives that a probability of the transition between initial ($i$) and final ($f$) states is proportional to
\begin{equation}\label{Probability1}
\mathscr{E}_0^2P_{fi}^2,
\end{equation}
where $P_{fi}$ is a non diagonal element of the polatization matrix from Eq.~\eqref{Polarization1}. Obviously, for the Q1D waveguide it is not zero only if the longitudinal parity of the states is the same whereas for the 3D geometry it vanishes if taken between the levels with different azimuthal numbers $m$. A physical quantity of practical importance is the oscillator strength
\begin{equation}\label{OscillatorStrength1}
f_{fi}=(E_f-E_i)P_{fi}^2,
\end{equation}
which contains again a square of $P_{fi}$. Also, the term 
\begin{equation}\label{Xi1}
\Xi=\omega P_{fi}^2\delta\!\left(E_f-E_i-\omega\right)
\end{equation}
enters into the expression of the linear optical absorption coefficient \cite{deSousa1,Tas1,Radu1} and photo ionization cross section.\cite{deSousa1} Oscillator strength $f_{20}$ is shown in Fig.~\ref{OscillatorStrengthQuasi1DFig1} for the Q1D waveguide with $s=4$. It is seen that as a function of the field it has a nonmonotonic behavior with a pronounced maximum at $\mathscr{E}_{max}\simeq1.27$ and the corresponding upper bound orbital disappears at $\mathscr{E}\simeq1.455$ when the excited-state waveform [see panel (d) of Fig.~\ref{FunctionsSecondExcitedFig1}] turns the oscillator strength to zero again, as it was for the flat geometry. Physically, the oscillator strength defines the probability of electron transition between the levels with the absorption or emission of one photon with its energy equal to the energy difference  $\Delta E$ between these orbitals. This dimensionless even in the regular units semi classical quantity measures the quantum mechanical emission or absorption rate in terms of its counterpart of the classical electron in the electromagnetic field when it can be deemed as a harmonic oscillator with eigen frequency $\Delta E$ perturbed by the time-periodic oscillations with frequency $\omega$.\cite{Thorne1} As Eqs.~\eqref{OscillatorStrength1} and \eqref{Polarization1} demonstrate, the magnitude of the oscillator strength is determined by, first, the overlap of the wave functions $\Psi_n$ and $\Psi_{n'}$ and, second, symmetry considerations. For our Q1D geometry, this latter requirement applied in the longitudinal direction eliminates, as ascertained above, the transitions between the states with different parity of their quantum numbers. Since at the zero field the waveforms are even functions of the variable $z$, $\left.\Psi_n(x,-z)\right|_{\mathscr{E}=0}=\left.\Psi_n(x,z)\right|_{\mathscr{E}=0}$, dipole transitions between the orbitals are forbidden too, $\left.f\right|_{\mathscr{E}=0}=0$. Applied voltage destroys mirror transverse symmetry leading to the growth of the oscillator strength. The overlap between  $\Psi_n$ and $\Psi_{n'}$ increases for the increasing electric intensity since both of them are shifted to the upper duct. However, this growth of the oscillator strength is terminated due to the rapid decrease of the absolute value of the excited-state wave function at the voltage quite close to the critical one. As panel (b) of Fig.~\ref{FunctionsSecondExcitedFig1} demonstrates, at $\mathscr{E}=1$ the extrema of $\Psi_n$ are almost unaffected by the field with its only influence being in the shift along the applied force what guarantees a sustainable growth of the oscillator strength in this regime. But, as the electric intensity approaches the critical one, the function starts to flatten, see, e.g., subplot (c) of Fig.~\ref{FunctionsSecondExcitedFig1} for $\mathscr{E}=1.4$. As a result, the oscillator strength reaches its maximum and for the higher voltages quite precipitously drops turning to zero at $\mathscr{E}_{cr}$.

\begin{figure}
\centering
\includegraphics[width=\columnwidth]{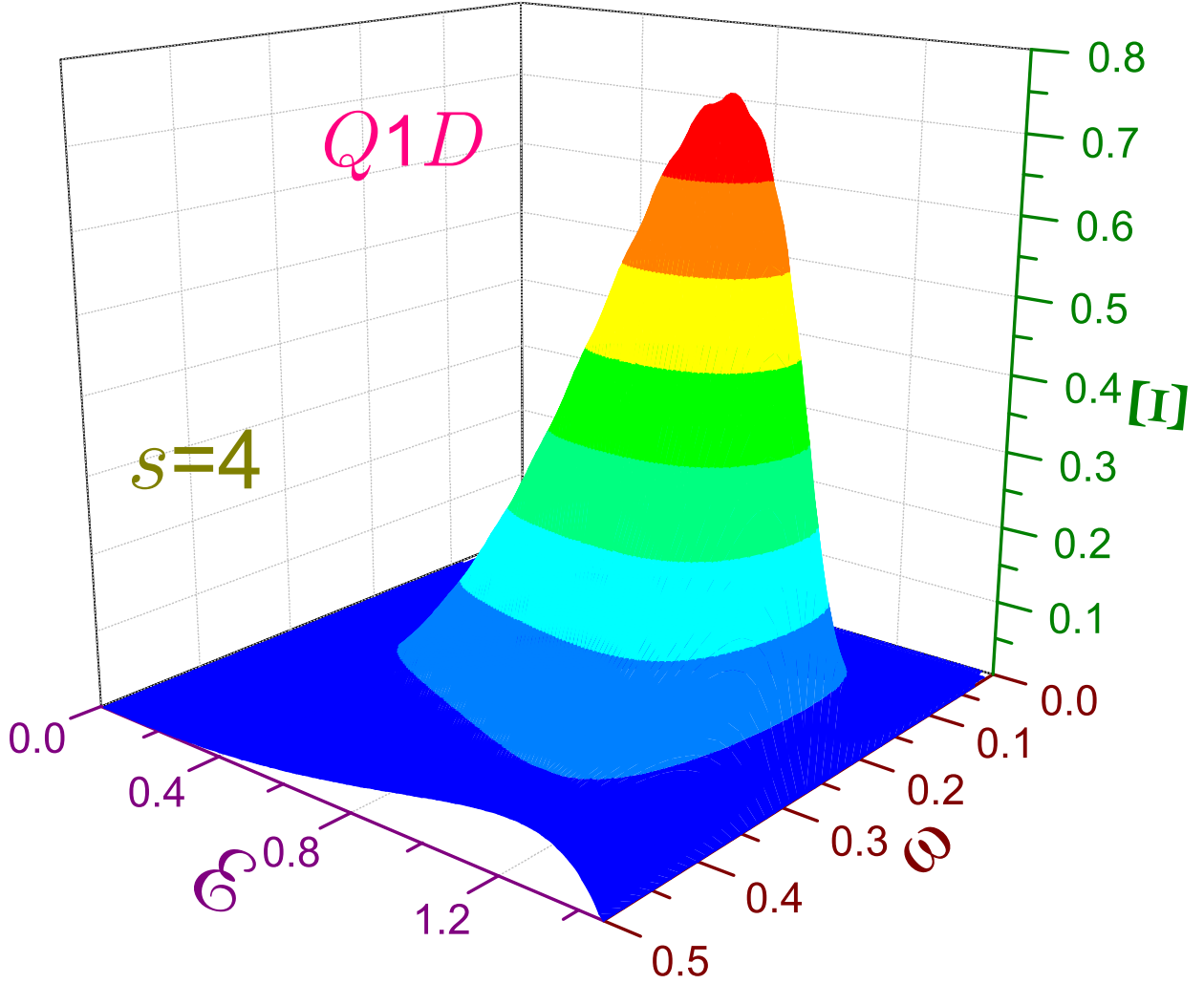}
\caption{\label{AbsorptionSpectrumFig1}
Quantity $\Xi$, Eq.~\eqref{Xi1}, as a function of the frequency $\omega$ and electric field $\mathscr{E}$ for the Q1D waveguide with $s=4$.}
\end{figure}

The energy difference $\Delta E$, which determines the frequency of the emitted or absorbed radiation, is depicted in Fig.~\ref{OscillatorStrengthQuasi1DFig1} by the dotted curve. It is seen that it is a monotonically decreasing function of the voltage.

\begin{figure}
\centering
\includegraphics[width=\columnwidth]{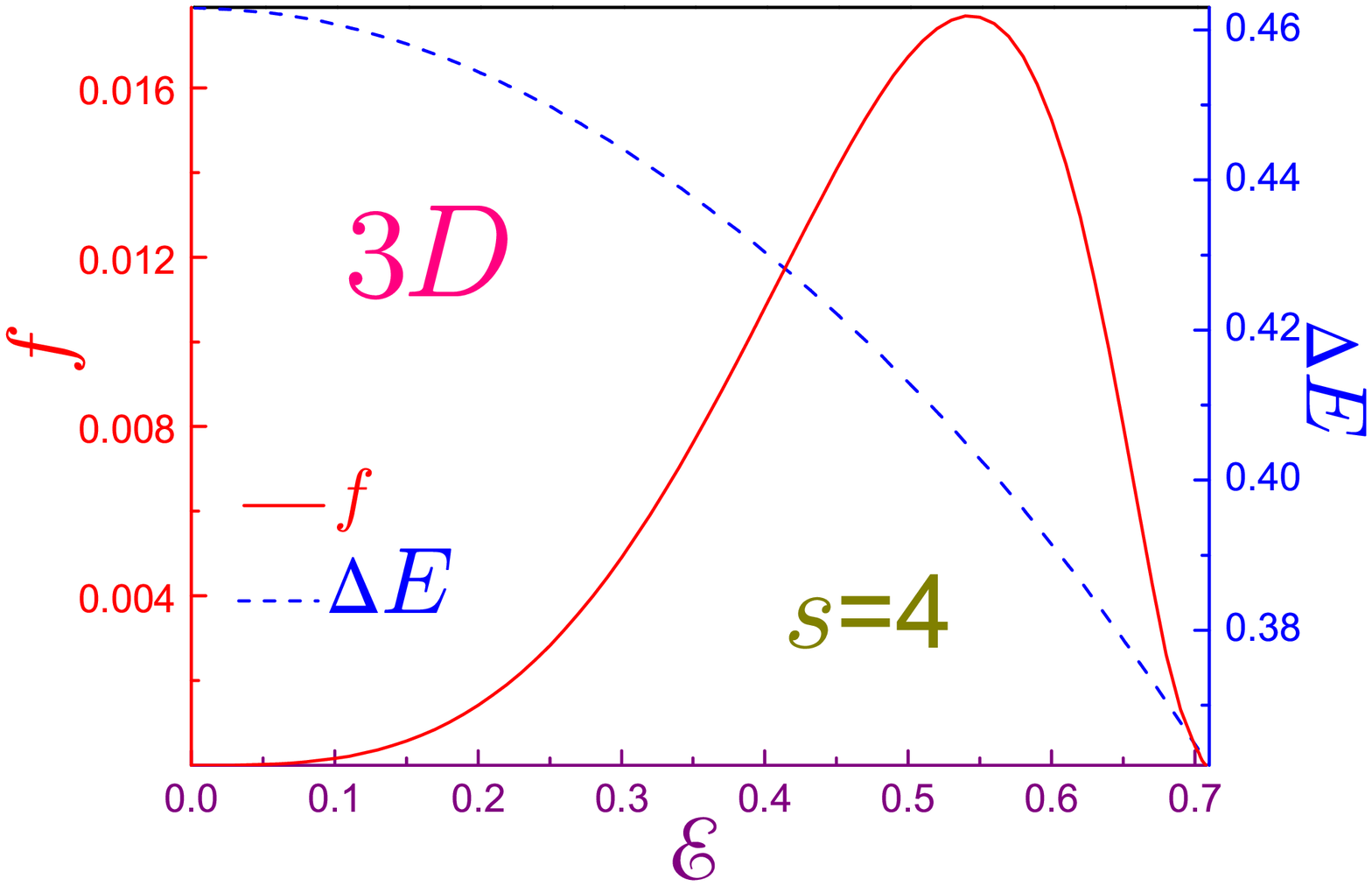}
\caption{\label{3DOscillatorStrengthFig1}
Oscillator strength $f$ (solid line, left axis) and energy difference $\Delta E$ for transition between the two $m=0$ states of the 3D waveguide with $s=4$ as functions of the electric field.}
\end{figure}

Dependence of the term $\Xi$ from Eq.~\eqref{Xi1} on the frequency $\omega$ and electric field $\mathscr{E}$ is shown in Fig.~\ref{AbsorptionSpectrumFig1} where, as usual, instead of the $\delta$-dependence, the Lorentzian
\begin{equation}\label{Lorentzian1}
\frac{1}{\pi}\frac{\Gamma_{fi}}{\left(E_f-E_i-\omega\right)^2+\Gamma_{fi}^2},
\end{equation}
has been used and the value of the half width $\Gamma_{fi}$ was taken to be $1/20$. It is seen that, similar to the oscillator strength, the shape of $\Xi$ is characterized by the pronounced maximum that is achieved at the same field as for the oscillator strength and at the frequency being equal to the energy difference at $\mathscr{E}_{max}$. Observe that at the critical field the structure becomes totally transparent for any frequency of the incident light.

\begin{figure}
\centering
\includegraphics[width=\columnwidth]{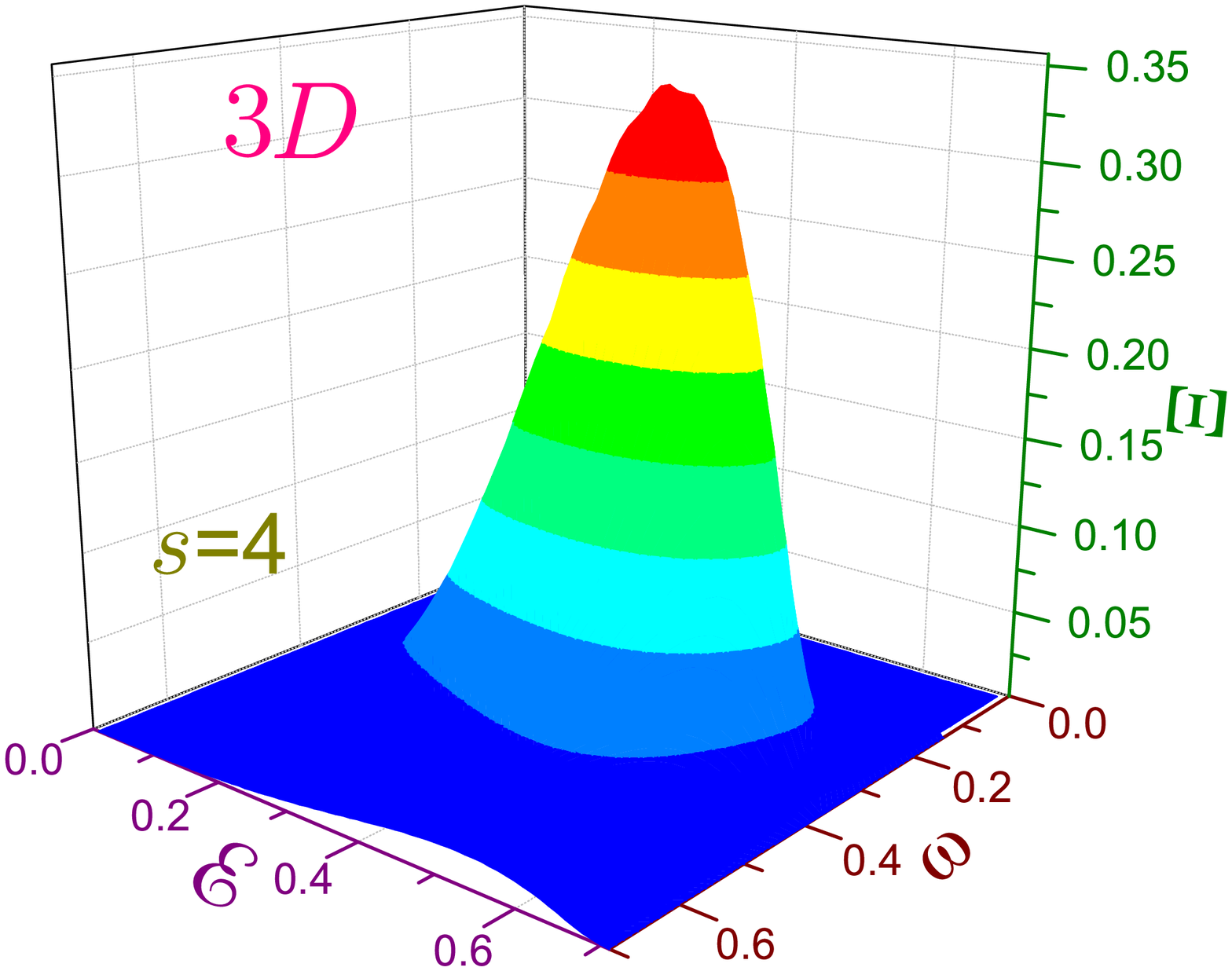}
\caption{\label{3DAbsorptionSpectrumFig1}
Quantity $\Xi$, Eq.~\eqref{Xi1}, as a function of the frequency $\omega$ and electric field $\mathscr{E}$ for the 3D waveguide with $s=4$.}
\end{figure}

Next, Figs.~\ref{3DOscillatorStrengthFig1} and \ref{3DAbsorptionSpectrumFig1} show the oscillator strength (together with the corresponding energy difference) and parameter $\Xi$, respectively, of the 3D coupled quantum duct with $s=4$ for the two lowest $m=0$ states. Qualitatively, they are identical with the Q1D case presented in Figs.~\ref{OscillatorStrengthQuasi1DFig1} and \ref{AbsorptionSpectrumFig1}. Quantitatively, for the 3D geometry smaller electric intensities are needed to destroy the excited bound orbital ($0.708$ vs. $1.455$) with the oscillator strength maximum being smaller than its Q1D counterpart ($0.0177$ vs. $0.0393$) whereas the corresponding energy difference is greater ($0.405$ vs. $0.288$). Due to the last fact, the $\Xi$ maximum on the frequency axis is achieved at the higher frequencies, as a comparison of Figs.~\ref{3DAbsorptionSpectrumFig1} and \ref{AbsorptionSpectrumFig1} reveals.

Finally, in Table ~\ref{Table1} the actual data for the GaAs structure with the electron effective mass $m_p=0.067m_e$ ($m_e$ is a free electron mass) are provided that were computed for $d_1=d_2=10$ nm and $s=40$ nm by transforming back to the regular units from the dimensionless ones introduced before Eq.~\eqref{Psi3DRhoLessA1}. It shows that for both geometries the critical fields are smaller than $10^5$ V/cm what can be easily achieved experimentally.\cite{Miller1,Miller2,Weiner1} Maximum of the electromagnetic absorption or emission lies either in the visible part of the spectrum (3D DQW) or in its nearest vicinity (Q1D configuration). Table~\ref{Table1} and Figs.~\ref{OscillatorStrengthQuasi1DFig1} -- \ref{3DAbsorptionSpectrumFig1} affirm that by applying the electric field one can efficiently change the optical properties of the system in a wide range.

\begin{table*}[ht]
\caption{ }
\begin{center}
    \begin{tabular}{||p{11cm}|c|c||}
    \hline
    Parameter&Q1D&3D\\ \hline
     Electric field at which upper bound state dissolves, $\times10^6$ V/m&8.11&3.95\\ \hline
    Electric field at which optical maxima are achieved, $\times10^6$ V/m&7.08&3.01\\ \hline
    Energy difference at the optical maximum, meV&16.2&22.8\\ \hline
    Corresponding wavelength, nm&768.6&543.5\\ \hline
    Oscillator strength at maximum&0.0393&0.0177\\
    \hline
    \end{tabular}
\end{center}
\label{Table1}
\end{table*}

\section{Concluding remarks}\label{sec_Conclusions}
Electric field influence on the quantum objects has been in the tight focus of the scientific interest since the early days of the wave mechanics. Despite the long history, an enthusiasm in pursuing this research is not fading. The main message of the present analysis promulgates that the transverse electric field applied to two window-coupled straight waveguides strongly modifies their electronic and optical properties; namely, the ground bound state survives any magnitude of the applied voltage whereas the existence of the excited localized orbitals can be efficiently controlled by the voltage what leads to nonmonotonic dependence of the optical properties such as the oscillator strength, linear absorption coefficient and photo ionization cross section, on the intensity $\mathscr{E}$.

To computationally simplify our discussion, the form of the potential $V({\bf r})$ from Eq.~\eqref{Schrodinger1} was chosen as depicted in Fig.~\ref{Schematics}. In this case, for the zero field, $\mathscr{E}=0$, the switching power of the CW is determined by its length $s$ only. Other frequently used shape of the common wall is a smooth Gaussian potential with a parabolic saddle-like dependence $\exp\left(-\alpha_x^2x^2-\alpha_z^2z^2\right)$ in the coupling region.\cite{Abdullah1,Abdullah2,Abdullah3} Then, the inter-wire tunneling is determined not only by the opening size, which is defined as $2/\alpha_x$, but by the barrier height too. Just this form of the potential was used to explain experimental data obtained from the split-gate structures.\cite{Ramamoorthy1,Ramamoorthy2} To tune the switching power of this device, the voltage was applied to the Cr/Au "finger" gates deposited on a GaAs/AlGaAs heterostructure, see Refs.~\onlinecite{Ramamoorthy1,Ramamoorthy2} for more information. The same technique can be employed to check the optical results derived above.

In our analysis, we considered optical transitions inside conduction band only. It is easy to extend it from the intraband to interband interaction; namely, if the electron from its lowest-energy orbital makes a transition to the hole ground level or the hole recombines with the electron both of which are in their ground state, the corresponding excitonic peak will survive any electric field (though its maximum will decrease since the overlap integral between the electron and hole waveforms is getting smaller at the higher $\mathscr{E}$ as they are pushed in the opposite directions) whereas if either (or both) of the excited levels is involved, the interband transition between them will be subdued and eventually eliminated by the applied voltage. One can also talk about the waveguides interaction with the circularly polarized light with its electric field vector lying in the $x-y$ plane. Then, for the 3D geometry, the expressions for the oscillator strength and linear absorption coefficient will include a square of the dipole transition element $T_{fi}$, which, however, reads now as
\begin{equation}\label{AbsoptionCircular1}
T_{fi}=\left\langle f\left|\rho e^{i\varphi}\right|i\right\rangle,
\end{equation}
what imposes a natural selection rule for the azimuthal quantum number, $\Delta m=\pm 1$. Since there are no analytic expressions for them in known literature,\cite{Gradshteyn1,Prudnikov2,Prudnikov3,Brychkov1} radial integrals now have to be evaluated numerically. Obviously, at the critical electric field when the upper level is on the brink of collapse with its waveform turning to zero, the structure again becomes transparent for any optical frequency. Thus, the influence of the variation of the applied transverse voltage allows to change the optical characteristics of the coupled waveguides what can be used in such devices of optonanoelectronics as sensors and filters with their frequency-dependent transmittance being regulated by the electric fields.

\acknowledgements
Research was supported by SEED Project No. 1702143045-P from the Research Funding Department, Vice Chancellor for Research and Graduate Studies, University of Sharjah.

\end{document}